\documentclass[preprint]{aastex}

\usepackage{pstricks}
\usepackage{graphicx}
\usepackage{txfonts}

\usepackage{natbib}

\newcommand{\ca}{\ion{Ca}{2} 8542 \AA}
\newcommand{\fe}{\ion{Fe}{1} 6302 \AA}
\newcommand{\arc}{$^{\prime\prime}$}
\newcommand{\added}[1]{{#1}}

\begin{document}

\title{Kink waves in an active region dynamic fibril }

\author{A.Pietarila}
\affil{National Solar Observatory, 950 N. Cherry Avenue, Tucson, AZ 85719
, USA}

\and

\author{R. Aznar Cuadrado, J. Hirzberger and
  S.~K. Solanki\altaffilmark{1}} \affil{ Max-Planck-Institute for
  Solar System Research, Max-Planck-Strasse 2, 37191
  Katlenburg-Lindau, Germany}

\altaffiltext{1}{School of Space Research, Kyung Hee University,
  Yongin, Gyeonggi, 446-701, Korea}

\date{}


\begin{abstract}
We present high spatial and temporal resolution \ca\ observations of a
kink wave in an on-disk chromospheric active region fibril. The
properties of the wave are similar to those observed in off-limb
spicules. From the observed phase and period of the wave we determine
a lower limit for the field strength in the chromospheric active
region fibril located at the edge of a sunspot to be a few hundred
Gauss. We find indications that the event was triggered by a
small-scale reconnection event higher up in the atmosphere.
\end{abstract}
\keywords{Sun: chromosphere, Sun: magnetic fields, Sun: sunspot, Sun:
  oscillations }

\section{Introduction}

Recent observations have strengthened the connection between spicules
and coronal heating, either via spicules supplying hot plasma to the
corona \citep{DePontieu+others2011} or energy dissipation from MHD
waves propagating along spicules
(e.g.,~\citealt{Kukhianidze+others2006,DePontieu+others2007,He+others2009}).
The first observations of transverse oscillations, indicative of MHD
waves, in spicules were made already in the late 1960s
\citep{Pasachoff+others1968}. Since the discovery that MHD waves are
fairly ubiquitous in spicules \citep{DePontieu+others2007} the
interest has been renewed. For a recent observational review of
spicule oscillations see \cite{Zaqarashvili+Erdelyi2009}. Several
authors, e.g., \cite{DePontieu+others2007,He+others2009}, have
identified the waves as kink or Alfv\'en and estimated that they carry
enough energy to heat the corona.

Based on theoretical considerations we expect to find four different
modes of waves in cylindrical plasma geometries appropriate for solar
conditions: kink, sausage, longitudinal, and torsional modes. The
first three modes are compressible, while the fourth one is not. All
of the modes have been observed in the solar corona (although the
identification of the pure, and thus incompressible, Alfv\'en waves
has been questioned by \citealt{VanDoorsselaere+others2008}). Alfv\'en
and kink waves can be triggered by granular buffeting of photospheric
magnetic flux tubes (e.g.,
\citealt{Roberts1979,Hollweg1981,Spruit1981,
Choudhuri+others1993,Huang+others1995,Musielak+Ulmschneider2001}) or
by small-scale reconnection
\citep{Axford+McKenzie1992}. Differentiating observationally between
kink and Alfv\'en wave modes is not entirely straightforward since
both can cause transversal displacements and observationally appear as
incompressible, at least for some magnetic field geometries
\citep{VanDoorsselaere+others2008}. In general, these wave modes can
be distinguished in straight, cylindrical, high density magnetic flux
tubes. Flux tubes act as wave guides for kink waves (i.e., the wave is
confined in the tube), which in turn cause transverse displacements of
the tube axis. In contrast, Alfv\'en waves propagating incompressibly
along a magnetic flux tube are torsional and do not lead to a
transversal displacement
\citep{Erdelyi+Fedun2007,VanDoorsselaere+others2008}. Chromospheric kink and
Alfv\'en waves have been observed in off-limb spicules. Since fibrils
are the likely on-disk counterpart of spicules (e.g.,
\citealt{Tsiropoula+others1994}) one would expect to see transverse
oscillations in them as well.

The observed properties of MHD waves can be used to measure the
magnetic field strength in, e.g., coronal loops (e.g.,
\citealt{Nakariakov+Ofman2001}) as well as spicules (e.g.,
\citealt{Zaqarashvili+others2007,Singh+Dwivedi2007,Kim+others2008}). The
thus measured spicule field strengths outside active regions are
generally in the 10-40 G range. Spectropolarimetric measurements of
spicules yield similar field strengths, e.g., 10 G from Hanle
measurements of the \ion{He}{1} 10830 \AA\ triplet
\citep{TrujilloBueno+others2005} and 30 G from \ion{He}{1} D3
measurements \citep{LopezAriste+Casini2005}. In this paper we present
observations of transverse oscillations in on-disk fibrils, and based
on them, give an estimate of the magnetic field strength in
chromospheric active region dynamic fibrils.

\section{Observations and data analysis}
 
The main goal of the observing sequence was to study the chromosphere
on small spatial and temporal scales using the \ca\ line.
The line wings sample the photosphere up to the reversed granulation,
while the line core is formed in the chromosphere. Because of its
sensitivity to velocities, temperatures, and magnetic fields, the
\ca\ line is a powerful tool for chromospheric studies.

We observed active region AR 11019 ($x=-342$\arc\ and $y=397$\arc,
$\mu$=0.87) with the CRisp Imaging SpectroPolarimeter (CRISP,
\citealt{crisp}) at the Swedish Solar Telescope on June 2, 2009 from
08:15 UT to 08:50 UT. The observing sequence, optimized for fast
chromospheric dynamics, consisted of a main sequence of rapid
Dopplergrams (at $\pm$ 180 m\AA\ from line core) of the \ca\ line with
a more thorough spectral scan (referred to as the background, BG,
scan) interwoven into it. Details are given in Table
\ref{table:obs}. In practice, we alternated between the two scans,
i.e., every second wavelength of the series was either the red or blue
point of the Dopplergram scan and the remaining points made up the
background scan. Every 5.3 minutes the \fe\ line was scanned to
provide a photospheric context. The transmission profile of the
instrument is 107.3 m\AA\ at 8540 \AA\ and 53.5 m\AA\ at 6300 \AA. The
pixel size is 0.0592\arc.  For each image recorded by CRISP a
wide-band (2 \AA) intensity image was simultaneously recorded with a
separate camera. These were later used to coalign the data set and to
destretch it to remove artifacts caused by the varying seeing
conditions following the method described by \cite{destretch}.

\begin{table}
  \caption{Observations}
  \label{table:obs}      
  \centering                          
  \begin{tabular}{lll}
    \hline\hline
    Scan & $\lambda-\lambda_{0}$ [m\AA]  & Cadence [s] \\
    \hline    
    \ca, Dopp & $\pm$ 180 & 3.4  \\
    \ca, BG & $\pm$ 750, $\pm$ 450, $\pm$ 350, & 21  \\
       & $\pm$ 250, $\pm$ 180, $\pm$ 90, 0, 2000 & \\
    \fe & $\pm$ 120, $\pm$ 60, 0, 680 & 320 \\
    \hline
  \end{tabular}
\end{table}

Adaptive optics (AO, \citealt{sstao}) were used during the
observations and the recorded images were reconstructed using
Multi-object multi-frame blind deconvolution (MOMFBD,
\citealt{momfbd}) resulting in images with very high spatial
resolution ($\approx$ 0.21 \arc, which corresponds basically to the
diffraction limit of the SST at 8540 \AA). For most of the time the
seeing conditions were good and stable, but they began to gradually
deteriorate toward the end of the observations.


For the spectropolarimetric data instrumental polarization effects of
the laboratory setup were measured with dedicated calibration
optics. Additionally, for the \fe\ data the telescope polarization was
determined using a model developed by \cite{Selbing2005}. No telescope
model is currently available for \ca\, so this step was omitted in the
\ca\ reduction. These demodulation matrices were applied to the data
resulting in the full Stokes vectors for the observed wavelength
points for each pixel. \footnote{Comparison of the \fe\ and \ca\
Stokes $V$ signals show that the \ca\ circular polarization data is
not severely affected by instrumental polarization effects and can be
used, at least, for purely qualitative purposes
(Fig.~\ref{fig:fov}). Because of the much smaller signals, the \ca\
linear polarization signal cannot be used without first applying a
proper telescope model.}


\section{Results}

\begin{figure*}
\includegraphics[width=10cm]{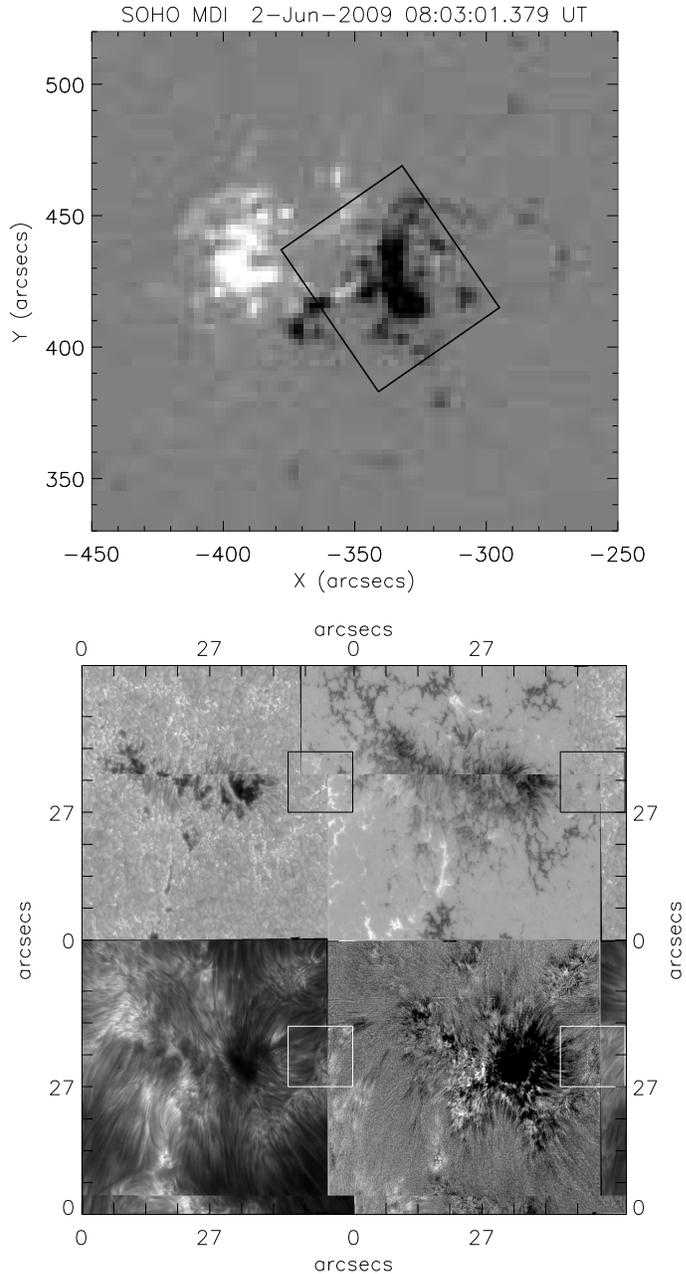}
\caption{Top frame: MDI magnetogram (color scale is $ -700
\mathrm{\mbox{\,--\,}} 700 \mathrm{\ G}$). The square marks the
approximate location of the SST FOV. Bottom frame: Field of view
observed by CRISP (images taken at 08:25 UT). Top left subframe:
wide-band (photospheric) intensity near \ca. Top right subframe:
Stokes $V$ at $-60$ m\AA\ from the \fe\ core. Bottom left subframe:
Intensity at $-90$ m\AA\ from the \ca\ core. Bottom right subframe:
Stokes $V$ at $-90$ m\AA\ from the \ca\ core. The squares in each
panel mark the location of the region of interest (ROI). Note that the
top frame is rotated $\approx$ 80 degrees counterclockwise compared to
the frames below.}
\label{fig:fov}
\end{figure*}

The main feature in the observed field of view (roughly 40 Mm $\times$
40 Mm) is AR 11019 (Fig.~\ref{fig:fov}). It consists of some pores and
a sunspot with only very few penumbral fibrils. The region of interest
(ROI, outlined in Fig.~\ref{fig:fov}) is on the edge of the main
spot. There are no large concentrations of magnetic field of either
polarity to the right of the observed region in the MDI magnetogram
taken 10 min before the observations. The following spot is located on
the opposite side of the observed main spot than the ROI.

A comparison of the \fe\ and \ca\ Stokes $V$ images in
Fig.~\ref{fig:fov} reveals a few locations, e.g., around position
(24,22) Mm and (21,16) Mm, where the two appear to have seemingly
opposite polarities. These may be a result of emission in the \ca\
line rather than a genuine change in polarity of the magnetic field as
pointed out by \cite{SanchezAlmeida1997}.

\subsection{Atmospheric structure and dynamics in the ROI}

\begin{figure*}
\includegraphics[width=18cm]{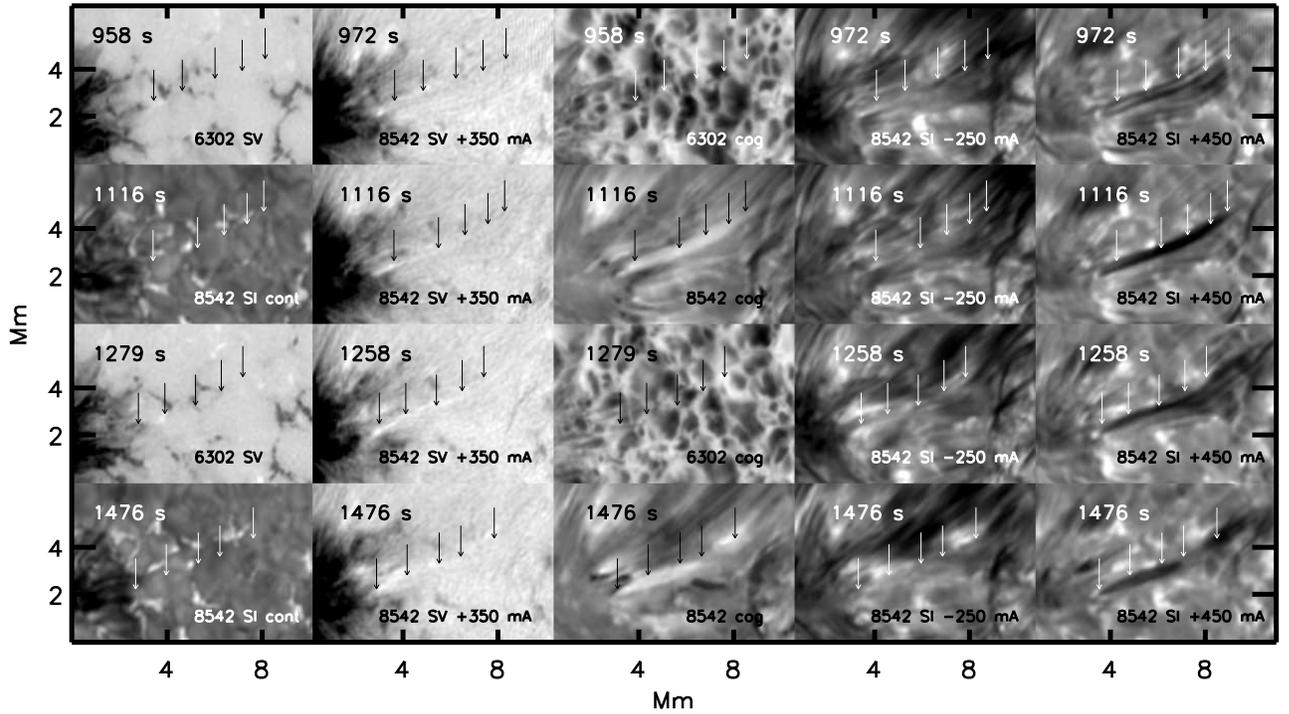}
\caption{ROI. $1^{st}$ column: alternating \fe\ Stokes $V$ signal and
  wide-band intensity near \ca. $2^{nd}$ column: \ca\ Stokes $V$ at
  +350 m\AA. $3^{rd}$ column: alternating \fe\ COG $\lambda$ (scaled
  to $\pm$1.9 km/s) and \ca\ COG $\lambda$ (scaled to $\pm$7
  km/s). $4^{th}$ column: \ca\ Stokes $I$ at -250 m\AA. $5^{th}$
  column: \ca\ Stokes $I$ at +450 m\AA. Observing time (in s from
  $t=0$) is given at the bottom of each image. The arrows point to the
  location of the FOI. }
\label{fig:xy}
\end{figure*}

The main chromospheric features in the ROI (Fig.~\ref{fig:xy}) are
fibrils originating from the spot and extending nearly radially away
from it. Most of the magnetic field to the right of the spot in and
around the ROI is unipolar. It is unlikely that many fibrils (assuming
they outline the magnetic field) in the ROI connect back to the
photosphere inside or near the observed area.

Dynamic fibrils and rapid blue-shifted events are seen in movies of
the \ca\ intensities. Dynamic, dark (when viewed in line wing
intensities) fibrils are seen relatively often. Usually they are
located some distance away from the spot and do not connect to it. An
exception is seen in image sequences of the red wing of \ca\ in
Fig.~\ref{fig:xy}: a collection of very dark fibrils connect to the
spot and are marked by the vertical arrows (to avoid confusion from
hereon this collection of fibrils will be referred to as FOI, fibrils
of interest). 2-3 individual fibrils make up the FOI which all
together are around 1 \arc in width. The FOI are visible in movies of
the wavelength points redward of the line core. When viewed in movies
of intensity at +750 m\AA\ the FOI start as faint fibrils which merge,
become darker and eventually retract into the spot. The footpoint
(discussed in more detail in section \ref{sec:fp}) of the FOI is
brighter than the surroundings and after the FOI connect with the
spot, the footpoint momentarily has positive polarity in \ca\
magnetograms. This does not necessarily imply a fleeting change in the
polarity of the magnetic field, but is more likely associated with an
inversion in the line core due to transient heating of the
chromosphere at this location \citep{SanchezAlmeida1997}.

\subsection{Periodic transverse displacement of FOI}

\begin{figure*}
\includegraphics[width=14cm]{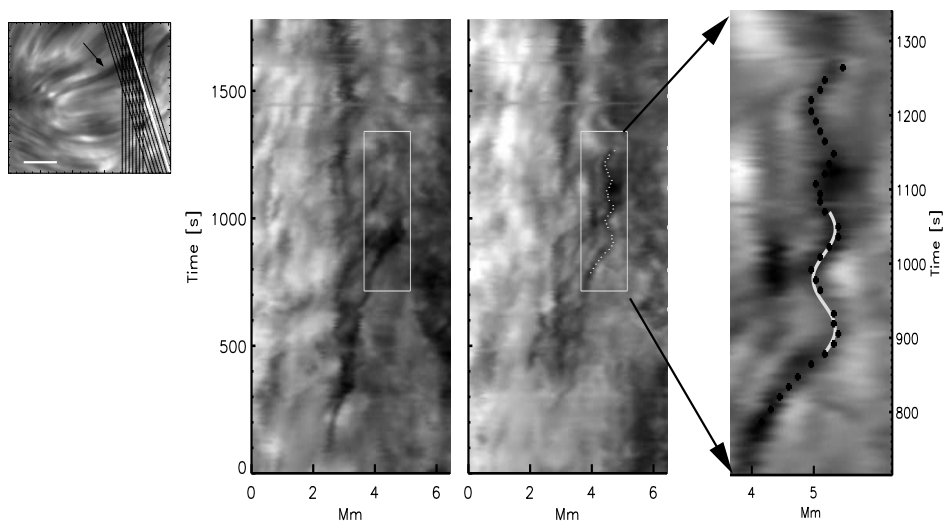}
\caption{Transverse oscillations of the FOI. Locations of the
  $xt$-cuts in the ROI (at +180 m\AA\ from line core) on left (white
  marks location of cuts shown in this figure, black shows location of
  all $xt$-cuts). Separate horizontal white line marks scale of 2
  Mm. Black arrow points to the FOI. Middle two panels: $xt$-cuts (at
  $-180$ m\AA\, left, and at $+180$ m\AA\, right, from \ca\ core).
  Right panel: magnification of area in the white rectangle. \added{Black crosses are the measured trajectory and the white line shows the sine-wave fitted to the trajectory.}    }
\label{fig:per2}
\end{figure*}

Fig.~\ref{fig:per2} shows $xt$(space-time)-cuts (artificial slit time
series) of the Dopplergrams placed perpendicular to the FOI. The FOI
start as a single dark feature moving toward the upper part of the ROI
(right in the $xt$-cut) with an apparent transverse speed of
approximately 2 km/s. The movement is at first best seen in the blue
wing and later in the red as well. At $t=300$ s the feature splits
into multiple strands which oscillate transversely. The oscillations
last $\approx$ 10 min after which the FOI disappear from the $xt$-cut,
\added{i.e., no damping in time of the oscillations is seen}. In the
ROI the FOI continue to retract and eventually (at the end of the
observing sequence) disappear entirely into the footpoint. The
coherent motions with a period of $\approx$20 s in the FOI are caused
by seeing conditions.

\begin{figure*}
\includegraphics[width=14cm]{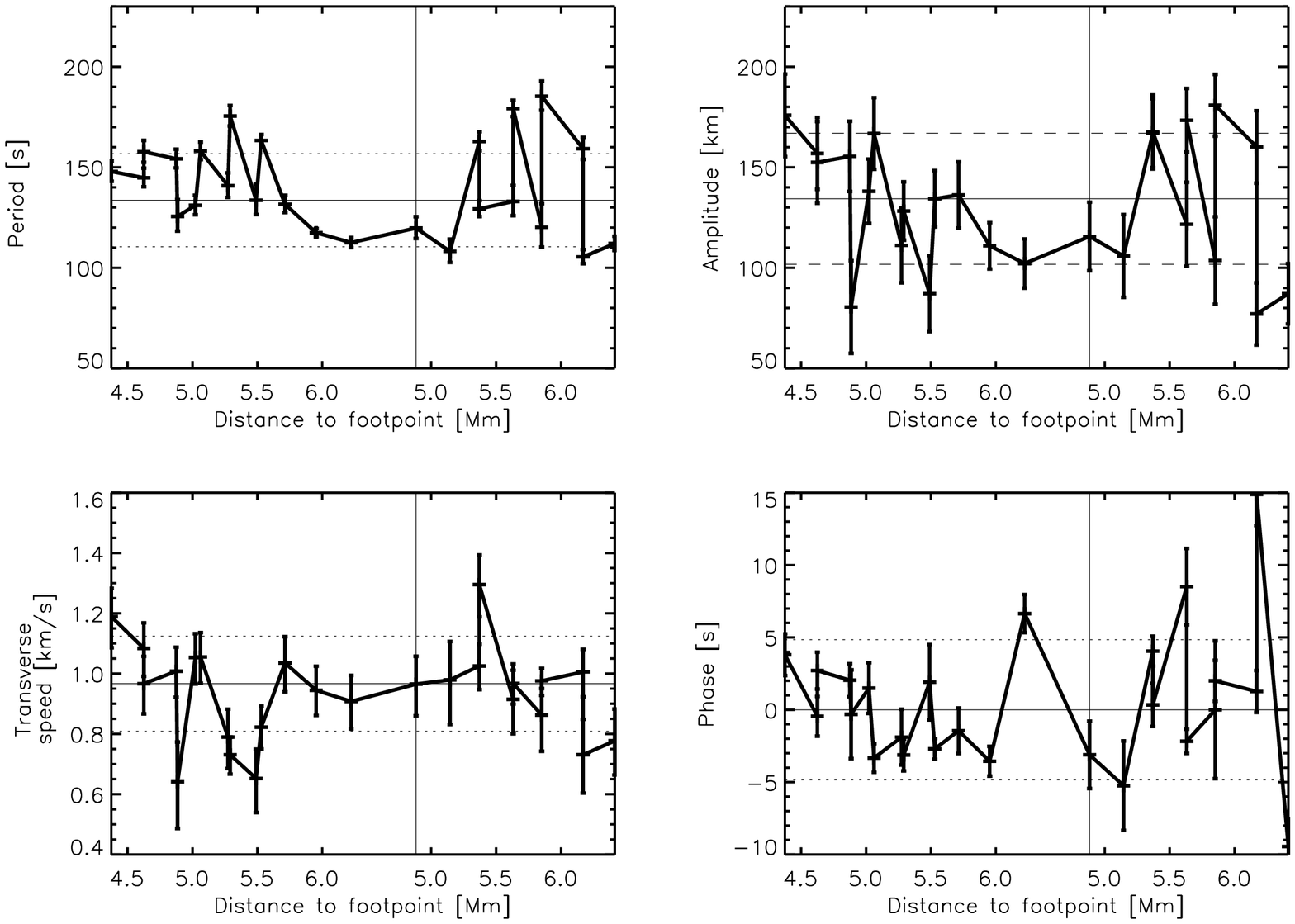}
\caption{Fits to transverse oscillations. Top left:
  period. \added{First 15 measurements are for slit positions
  perpendicular to the FOI and the rest for vertical slits (see Fig.~\ref{fig:per2}) }. Error
  bars show the standard deviation of the fits. Top right:
  amplitude. Bottom left: transverse velocity, i.e., apparent motion
  along the slit. Bottom right: phase. Phase=0 is the median phase of
  all measured oscillations. }
\label{fig:per1}
\end{figure*}

Measurements of the fibrils' trajectories in the red wavelength point
are made from $xt$-cuts placed perpendicular to the FOI or along the
vertical direction (no interpolation of data was made) at various
positions along the feature (positions of slits are shown in leftmost
panel of Fig.~\ref{fig:per2}). \added{The trajectories are identified
by eye from each $xt$-cut.} When a clear periodicity is visible in the
trajectory, a single sine wave \added{and a linear function that
accounts for any lateral movement of the FOI} are fitted to \added{the
trajectory}. The fit returns a period, amplitude, and phase relative
to $t=0$. A transverse velocity (amplitude/period) for the swaying
motion is also computed. The results of the fits are shown in
Fig.~\ref{fig:per1}. The period varies between 2-3 min, with a mean
value of 135 s. The mean apparent speed is 1 km/s. \added{The phase and distance to
the footpoint are not strongly correlated (Pearson correlation coefficient, $r_{c}$, is -0.26): the propagation is either too fast to be measured robustly on
such small spatial scales or the perturbation does not
propagate. The perpendicular to FOI phase measurements show, however, a more consistent decreasing trend in the interval [4.4-5.7 Mm]. An
ordinary linear least squares bisector fit to the phase as a function of
distance to the footpoint gives a slope of -5.26 $\pm$ 1.11 s/Mm
giving a phase speed of -190 km/s. (Note that the inferred slope depends on
which method is used to calculate the linear regression coefficients giving rise to a larger uncertainty than quoted above.) This, together with the negative correlation coefficient indicate that if the perturbation propagates, the propagation
is towards the footpoint.}

\begin{figure*}
\includegraphics[width=14cm]{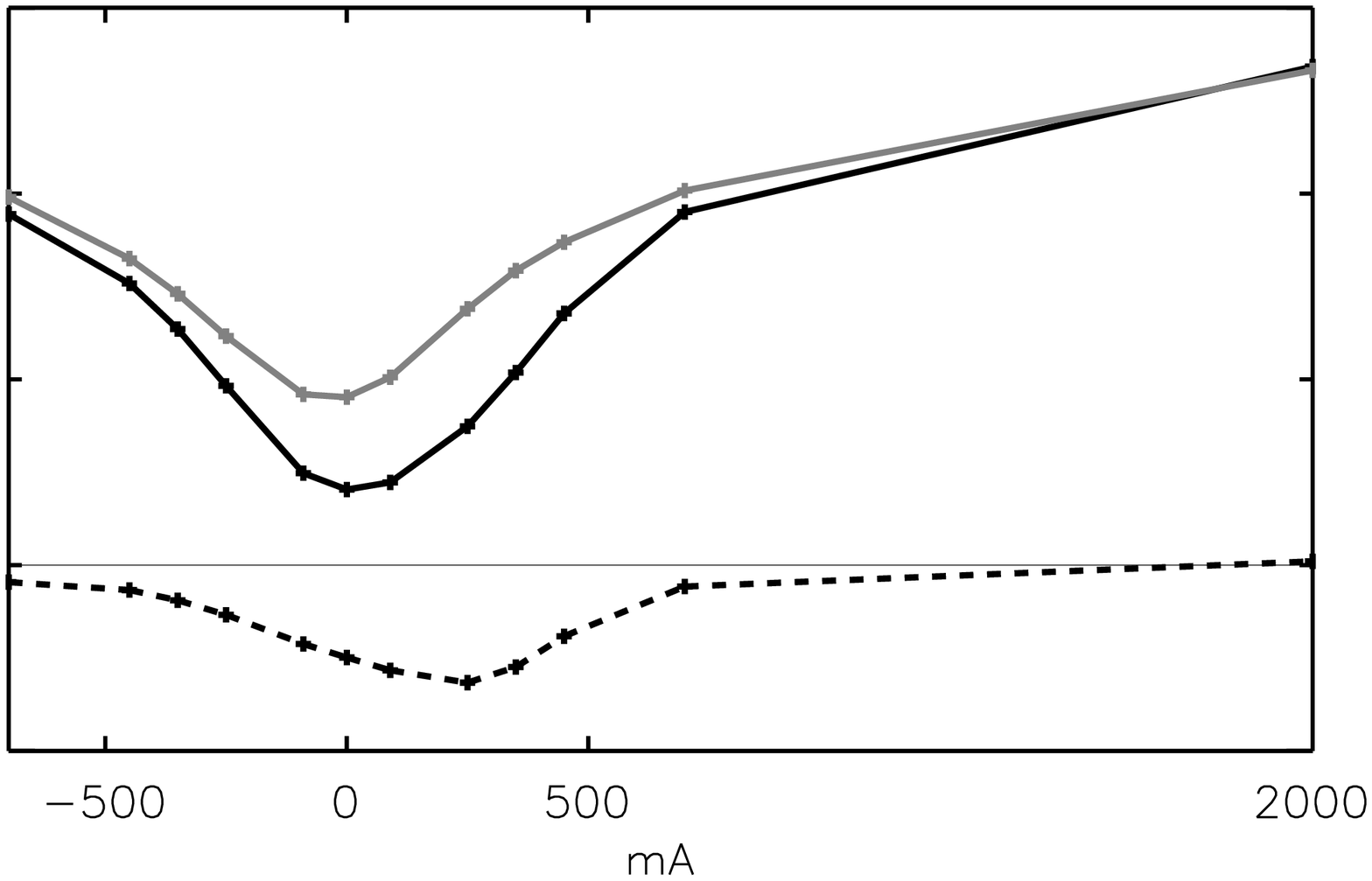}
\caption{Average \ca\ profile in the FOI (black solid line) and
  average line profile of the ROI excluding the FOI and spot
  (gray). The difference between the two upper profiles is indicated
  by the dashed black line. Y-axis units are arbitrary. Profiles are
  normalized to the local continuum.}
\label{fig:velstat}
\end{figure*}

We use the Dopplergrams to identify the locations in space and time
and interpolate them to the background scan to identify the FOI
pixels. Compared to the plage-dominated surroundings, the Stokes $I$
profiles in the FOI are deeper, red-shifted and more asymmetric, i.e.,
the red wing is broader than the blue wing
(Fig.~\ref{fig:velstat}). The difference between the FOI and
surrounding intensity profiles is largest at +250 m\AA. This is also
seen in measurements of the full width at half maximum (FWHM) and
center of gravity wavelength (COG $\lambda$). The median FWHM and COG
$\lambda$ in the FOI are 770 m\AA\ and 32 m\AA\ (corresponds to 1.4
km/s) compared to 615 m\AA\ and -17 m\AA\ (-0.6 km/s) in the
surrounding ROI (excluding the spot). (Note that zero wavelength and
velocity are not calibrated, so the COG $\lambda$ values are only
relative.) The FOI pixels closest to the footpoint tend to have
largest FWHM and COG $\lambda$; perhaps the material is accelerated as
it retracts to the footpoint or the FOI is more vertical near the
footpoint.

\subsection{Magnetic field strength in FOI}

The expression for kink waves in a vertical thin flux tube
\citep{Nakariakov+Verwichte2005} gives a relation between the magnetic
field strength in the flux tube and the wave period:

$$B_{0}=\sqrt{\frac{\mu_{0}}{2}}\frac{L}{P}\sqrt{\rho_{0}\left(1+\frac{\rho_{e}}{\rho_{0}}\right)}\rm{,}$$
where $\mu_{0}$ is the magnetic permeability, $\rho_{0}$ and
$\rho_{e}$ are the internal and external densities, $L$ is the
wavelength, and $P$ is the period.

\added{Since it is not possible to determine a reliable phase speed from the measurements, it is difficult to compute the
wavelength. We can, however, estimate a minimum value for $L$: the
oscillations are seen in a segment of approximately 1 Mm of length and
the uncertainty in the phase is of the order of the cadence (3.4 s)
leading to a minimum phase speed of 294 km/s and a wavelength of 40
Mm. (Note that this estimation gives no information on the propagation direction.) This $L$ is in agreement with estimates of wavelengths of kink waves
in spicules \citep{Kim+others2008}. For the density we use $1.4\times
10^{-11}$ g/cm$^{3}$. This value was used as a density at $z=1.25$ Mm
by \cite{Judge+Carlsson2010} to model spicules seen in \ion{Ca}{2}
H. They adopted the value based on what is a reasonable field
strength in the chromosphere (well below kG) and the highly supersonic
speeds observed in spicules. We will use 0.1 as the ratio of external
to internal densities. (The result does not depend strongly on the
ratio.) Based on these numbers and the observed periods, we estimate
the fibril magnetic field strength to be $ 220 \mathrm{\mbox{\,--\,}}
330 \mathrm{\ G}$. The mean period gives a field strength of 290
G. The corresponding Alfv\'en speeds are $ 165 \mathrm{\mbox{\,--\,}} 250 \mathrm{\
km/s}$. The inferred fibril field strengths are reasonable given the location of the fibril
emerging from the sunspot. Note that the wavelength, $L$, is the minimum wavelength based on uncertainty estimates and $B_{0}$ scales with
$L$. If we instead use the inferred, but highly uncertain, phase speed, 190km/s ($L$= 25 Mm), the resulting field
strengths are somewhat lower, $ 140 \mathrm{\mbox{\,--\,}} 210
\mathrm{\ G}$. (Corresponding Alfv\'en speeds are $ 100
\mathrm{\mbox{\,--\,}} 160 \mathrm{\ km/s}$). An additional large
uncertainty in the field strength arises from the lack of independent
density measurements in fibrils.}

\added{Note that the equation used to determine $B_0$ has been borrowed from coronal studies. It assumes a cylindrical geometry, zero plasma-$\beta$ and no flows. Approximating the chromospheric fibril geometry as cylindrical is reasonable, although the structures have some curvature. \cite{VanDoorsselaere+others2009} studied the effect of loop curvature on coronal loop kink oscillations and found that the kink wave period and damping are similar in straight cylinders with a density transition to the corona and in curved toroidal loops surrounded by a smooth transition to the corona. For a field strength of 290 G plasma-$\beta$ in the mid-chromosphere is below 0.01 (based on the \cite{Maltby+others1986} quiet Sun model atmosphere). Hence the plasma-$\beta$ condition is well fulfilled. Finally, the downflow speed of roughly 10 km/s is roughly an order of magnitude smaller compared to the estimated Alfv\'en speed, so that the results obtained from the employed equation are consistent with the conditions under which the equation is valid.}

\subsection{Footpoint}
\label{sec:fp}

\begin{figure*}
\includegraphics[width=12cm]{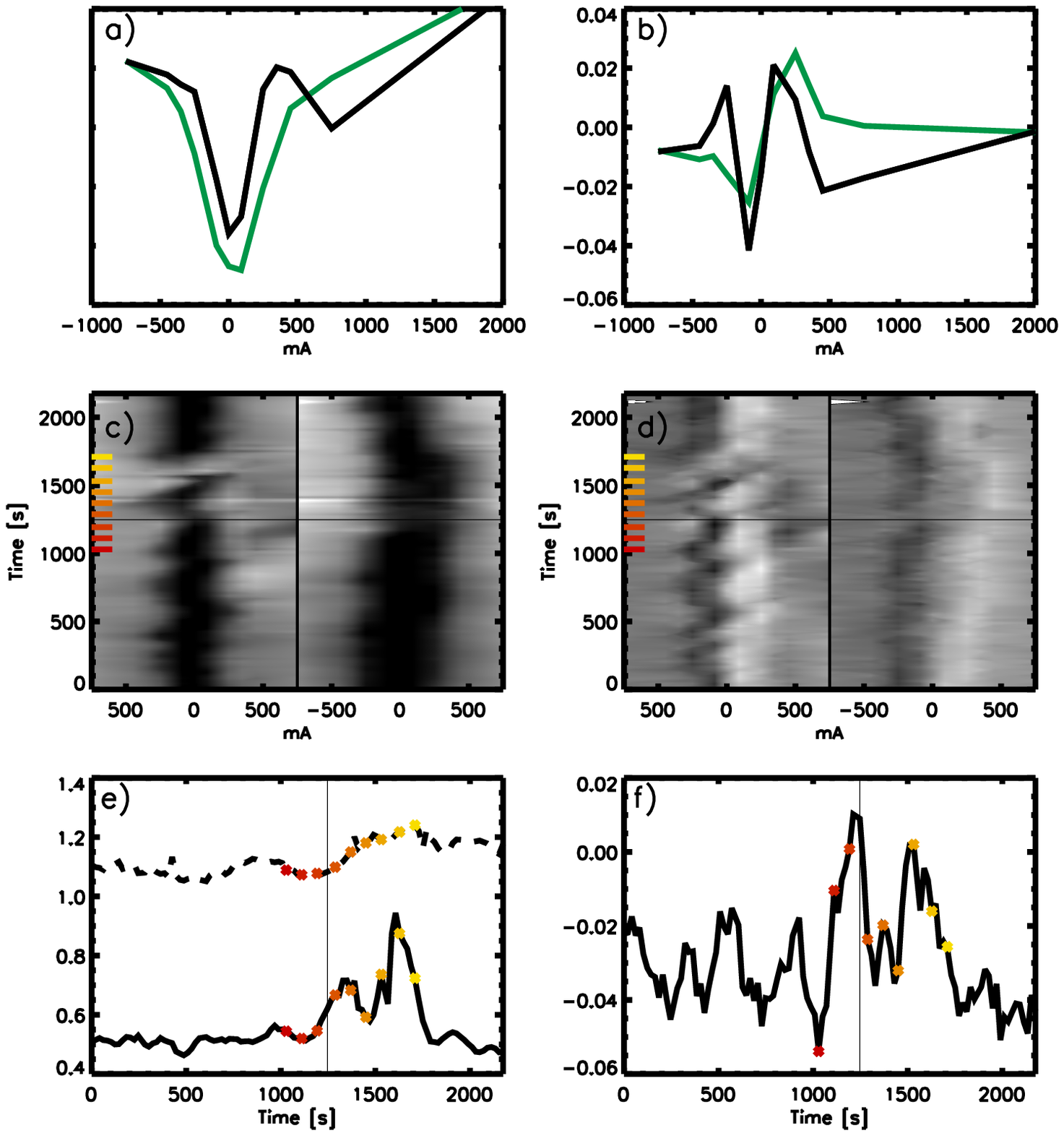}
\includegraphics[width=4cm]{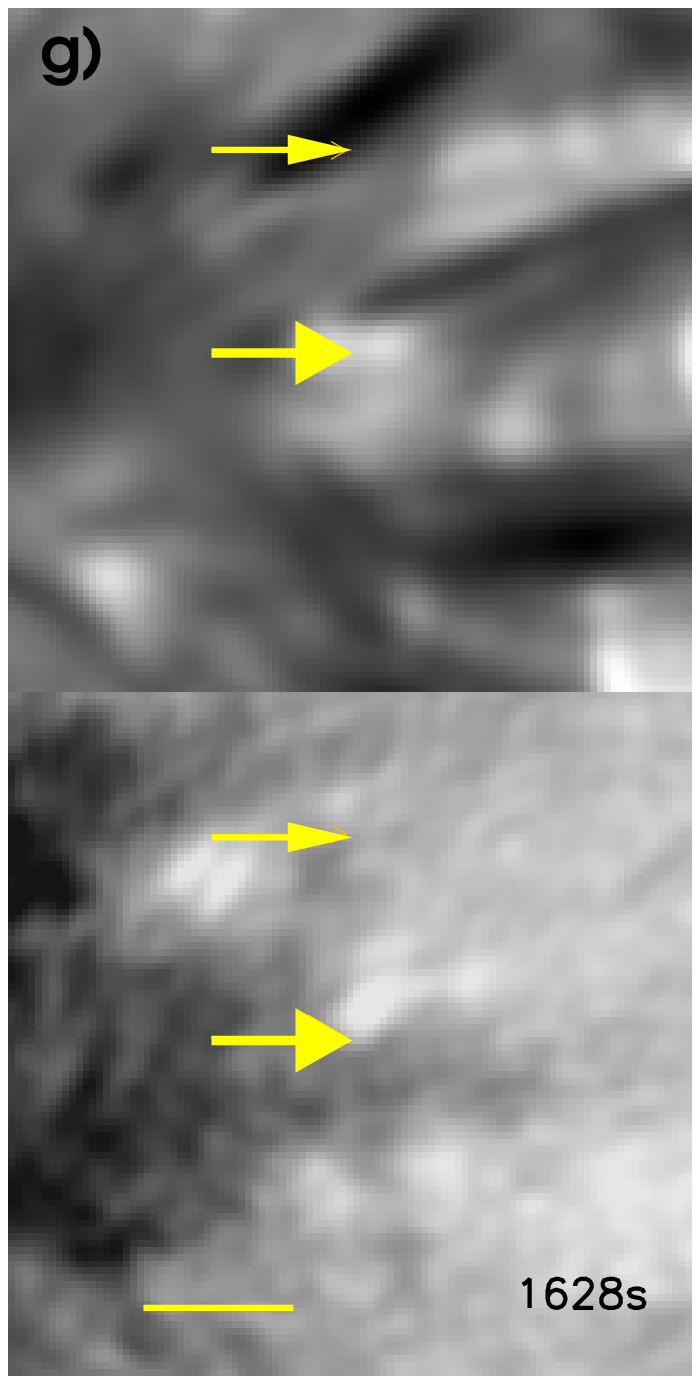}
\caption{FOI footpoint in \ca. Top panels: Stokes $I$ (panel a) and $V$
  (panel b) profiles in the footpoint (in black) at $t=1218$ s. The
  green (gray in the printed version) profiles are an average of a
  2.2\arc $\times$ 2.2\arc region around the footpoint. Middle panels:
  Time evolution of the Stokes $I$ (panel c) and $V$ (panel d)
  profiles in the footpoint (leftmost profile) and in a nearby pixel
  (rightmost profile). The horizontal line marks time of top panels'
  profiles, the colored ticks correspond to the times indicated by the
  asterisks in the lower panels. Bottom panels: Line core (solid line)
  and continuum (dashed line) intensities (panel e) and Stokes $V$ at
  -250 m\AA\ (panel f) in the footpoint. The vertical line shows the
  time of the profiles in the top panels. Panel g: snapshots of Stokes
  $I$ at 450 m\AA\ (top) and Stokes $V$ at -250 m\AA\ (bottom). The
  large (lower) arrows mark the location of the footpoint and small
  (upper) arrows the pixel shown in the middle panels. The tick mark
  corresponds to 1 Mm.}
\label{fig:footpoint}
\end{figure*}

The FOI footpoint is at the edge of the sunspot umbra on the side with
no penumbral fibrils.  The Stokes $I$ profiles are often asymmetric at
the footpoint. An example profile and the time evolution of the \ca\
line in the footpoint (i.e., where the FOI retract to) are shown in
Fig.~\ref{fig:footpoint}. Some of the features in the intensity
profiles may be due to the non-cotemporal wavelength sampling; there
is a 20 s time delay between the blue-most and red-most
wavelengths. The footpoint intensity time series displays a sawtooth
pattern indicating steepening waves (panel c in
Fig.~\ref{fig:footpoint}). The sawtooth pattern begins at t=300-400 s
when also the FOI first become clearly visible. Sawtooth patterns as
strong as in the footpoint are not usually seen in the surroundings. A
more diffuse sawtooth pattern is visible in the intensity time series
sampling the body of the FOI. The FOI pattern may be more diffuse
because the profiles originate from the fibrils as well as the
atmosphere the FOI are embedded in.

The footpoint line core and continuum intensities increase once the
dark fibril connects with the footpoint at t=1000 s. This coincides in
time with when the mostly negative Stokes $V$ blue lobe becomes
smaller in amplitude and even momentarily positive. No clear opposite
polarity flux is seen in the pore or right next to it at any point in
the \fe\ data (at much lower cadence) indicating that the changes in
\ca\ may be caused by a transient heating event in the
chromosphere. The footpoint intensity and Stokes $V$ time series
appear to have a 300 s periodicity, i.e., roughly twice the period of
the kink wave seen in the body of the FOI.

Note that the most dynamic phase, including the apparent opposite
polarity, begins at around t=1000 s when the FOI come into contact
with the footpoint, i.e., after the oscillations in the FOI have
already begun (at t=800 s); the rapid change in footpoint dynamics
appears to be a consequence of the retracting fibril.

\section{Discussion}

We have presented on-disk observations of a kink wave in an active
region chromospheric fibril. The high spatial and temporal resolution
of the data have allowed us to study the event in detail. It begins
with a lateral movement of the fibril which has its footpoint at the
edge of a sunspot umbra.  The transverse oscillations of the fibril
are associated with red-shifted, broadened intensity profiles, while
indications of \added{steepening waves} are seen in the footpoint. The
period of the waves in the footpoint is roughly twice the period of the FOI transverse
oscillations. The event ends as the fibril is retracted into the
footpoint. It is not clear what triggers the event. One possible
scenario is that the magnetic field above (at the base of the corona)
changes and both the lateral movement and oscillations result from
that. Since we see no \added{strong} systematic differences in
measured phases in our observations, we cannot say \added{for sure}
whether the disturbance propagates or, if it does, in which
direction. \added{Where a trend in measured phases is seen the
propagation is downwards, i.e., towards the FOI footpoint.} Based on a
comparison of Hinode observations and Monte Carlo simulations of
Alfv\'en waves in spicules, both upward and downward propagating waves
are likely \citep{DePontieu+others2007}.

If the observed event is due to reconnection higher up, the observed
red-shift would be consistent with a bipolar jet occurring at the
reconnection site (e.g., \citealt{Innes+others1997}). The largest
difference between the FOI intensity profile and the average intensity
profile of the surroundings is at $\approx$ 250 m\AA\, corresponding
to a Doppler velocity of 9 km/s (Fig.~\ref{fig:velstat}). Compared to
velocities of -20 km/s seen in the \ion{Ca}{2} 8542 \AA\ on-disk
counterparts of type II spicules (rapid blue-shifted events,
\citealt{RouppevanderVoort+others2009}), the FOI velocity does not
seem very high. \cite{Shoji+Kurokawa1995} observed downflows of up to
50-100 km/s in H-$\alpha$ and Ca K during the impulsive phases of two
flares. They also observed \ion{Na}{1} D1 and D2 lines as well as
metallic lines which showed significantly smaller, up to 2-6 km/s,
downflows. Based on the above, the magnitude of the FOI red-shift does
not exclude a small-scale reconnection event higher up in the
atmosphere as the trigger of the observed dynamics (no flares were
recorded during the observation day, so it can only be a small event).
A scenario involving a small scale reconnection event is also
supported by the apparent change in the \ca\ Stokes $V$ polarity in
the footpoint and the increased line and continuum intensities when
the FOI connect to the footpoint. These can be interpreted as an
emission feature near the line core caused by transient heating in the
chromosphere \citep{SanchezAlmeida1997}.

Fast and slow MHD waves may be coupled due to effects of boundary
conditions, inhomogeneities, and gas pressure. Coupling of MHD waves
is a possible explanation for the combined EIT and TRACE observations
of loop footpoint intensity variations having the same period as the
co-occuring fast transverse loop oscillation
\citep{Verwichte+others2010}. This was demonstrated by
\cite{Terradas+others2011}, who show how transverse motions of a
coronal loop can produce slow waves with the same periodicity as the
transverse oscillation. A coupling of MHD waves may be the connection
between the observed FOI oscillations and the \added{steepening}
waves. Note that unlike in the coronal loop case, the observed
\added{steepening (indicating the presence of shocks)} wave has a
period roughly twice that of the FOI oscillation.
 
The mean period, amplitude, and phase speed of the oscillations are
similar to measurements in off-limb spicules (e.g.,
\citealt{He+others2009} at heights below 2 Mm). The properties of the
oscillations allowed us to estimate the field strength in the dynamic
fibril to be at least $220 \mathrm{\mbox{\,--\,}} 330 \mathrm{\
G}$. For comparison, a value of 120 G at the base of the corona was
used by \cite{DePontieu+others2004} to model spicules via leakage of
P-modes into the chromosphere. \cite{Centeno+others2010} estimate,
based on \ion{He}{1} 10830 \AA\ triplet Hanle and Zeeman measurements,
50 G as a possible lower value for the field in network
spicules. \cite{LopezAriste+Casini2005} measured 30 G fields at a
height of $3000 \mathrm{\mbox{\,--\,}} 5000 \mathrm{\ km}$ above the
limb. Considering that we observed an active region and the magnetic
field strength decreases with height (the \ca\ on-disk observations
are at heights below 1500 km) a field strength of a few hundred Gauss
is reasonable. It is also consistent with the chromospheric field
strength near the outer boundary of a sunspot (e.g., \citealt{
Solanki+others2003,OrozcoSuarez+others2005}). Once a telescope model
for the \ca\ line becomes available, it will be interesting to compare
this value with results from Zeeman measurements.

How common kink waves in fibrils are is difficult to estimate, though
the similarity with kink waves in spicules suggests that the two
should have similar occurrence rates. We see indications of a few more
such oscillations in the data, but they are not as clear as the one
presented here. For the oscillation to be visible the spatial and
temporal resolutions need to be high. Also, the choice of wavelength
may play a role: perhaps the oscillations were so well visible because
the fibril was strongly red-shifted with respect to the
surroundings. Due to the complex optically thick background consisting
of other fibrils, the visibility of the oscillations on the disk is
limited compared to the optically thin limb spicules.

\section{Summary}

We identified kink waves in an active region fibril in high spatial
and temporal resolution \ca\ data.  The characteristics agree with
observations of kink waves in off-limb spicules. The properties, e.g.,
periods and velocity amplitudes, measured by \cite{He+others2009} for
kink waves seen in Hinode SOT Ca H imaging data at heights below 2 Mm
(relative to the reconnection site) are similar to those found in our
on-disk observations. The momentarily reverse polarity in \ca\ Stokes
$V$ profiles together with increased intensity point to a small-scale
reconnection event higher in the atmosphere as the trigger. To our
knowledge this is the first observation of chromospheric kink waves on
the solar disk. Based on the measured wave phase and period we
estimate the dynamic fibril field strength to be a few hundred Gauss.

\acknowledgements{This work was partly supported by the WCU grant
  No. R31-10016 from the Korean Ministry of Education, Science and
  Technology. The Swedish Solar Telescope (SST) is operated on the
  island of La Palma by the Institute for Solar Physics of the Royal
  Swedish Academy of Sciences in the Spanish Observatorio del Roque de
  los Muchachos of the Instituto de Astrof\'\i sica de Canarias. We
  thank the telescope staff for their kind support with the SST.}


\begin{thebibliography}{39}
\expandafter\ifx\csname natexlab\endcsname\relax\def\natexlab#1{#1}\fi

\bibitem[{{Axford} \& {McKenzie}(1992)}]{Axford+McKenzie1992}
{Axford}, W.~I. \& {McKenzie}, J.~F. 1992, in Solar Wind Seven Colloquium,
  Oxford: Pergamon Press, ed. {E.~Marsch, \& R.~Schwenn}, 1--5

\bibitem[{{Centeno} {et~al.}(2010){Centeno}, {Trujillo Bueno}, \& {Asensio
  Ramos}}]{Centeno+others2010}
{Centeno}, R., {Trujillo Bueno}, J., \& {Asensio Ramos}, A. 2010, \apj, 708,
  1579

\bibitem[{{Choudhuri} {et~al.}(1993){Choudhuri}, {Auffret}, \&
  {Priest}}]{Choudhuri+others1993}
{Choudhuri}, A.~R., {Auffret}, H., \& {Priest}, E.~R. 1993, \solphys, 143, 49

\bibitem[{{De Pontieu} {et~al.}(2004){De Pontieu}, {Erd{\'e}lyi}, \&
  {James}}]{DePontieu+others2004}
{De Pontieu}, B., {Erd{\'e}lyi}, R., \& {James}, S.~P. 2004, \nat, 430, 536

\bibitem[{{De Pontieu} {et~al.}(2011){De Pontieu}, {McIntosh}, {Carlsson},
  {Hansteen}, {Tarbell}, {Boerner}, {Martinez-Sykora}, {Schrijver}, \&
  {Title}}]{DePontieu+others2011}
{De Pontieu}, B., {McIntosh}, S.~W., {Carlsson}, M., {et~al.} 2011, Science,
  331, 55

\bibitem[{{De Pontieu} {et~al.}(2007){De Pontieu}, {McIntosh}, {Carlsson},
  {Hansteen}, {Tarbell}, {Schrijver}, {Title}, {Shine}, {Tsuneta}, {Katsukawa},
  {Ichimoto}, {Suematsu}, {Shimizu}, \& {Nagata}}]{DePontieu+others2007}
{De Pontieu}, B., {McIntosh}, S.~W., {Carlsson}, M., {et~al.} 2007, Science,
  318, 1574

\bibitem[{{Erd{\'e}lyi} \& {Fedun}(2007)}]{Erdelyi+Fedun2007}
{Erd{\'e}lyi}, R. \& {Fedun}, V. 2007, Science, 318, 1572

\bibitem[{{He} {et~al.}(2009){He}, {Marsch}, {Tu}, \& {Tian}}]{He+others2009}
{He}, J., {Marsch}, E., {Tu}, C., \& {Tian}, H. 2009, \apjl, 705, L217

\bibitem[{{Hollweg}(1981)}]{Hollweg1981}
{Hollweg}, J.~V. 1981, \solphys, 70, 25

\bibitem[{{Huang} {et~al.}(1995){Huang}, {Musielak}, \&
  {Ulmschneider}}]{Huang+others1995}
{Huang}, P., {Musielak}, Z.~E., \& {Ulmschneider}, P. 1995, \aap, 297, 579

\bibitem[{{Innes} {et~al.}(1997){Innes}, {Inhester}, {Axford}, \&
  {Wilhelm}}]{Innes+others1997}
{Innes}, D.~E., {Inhester}, B., {Axford}, W.~I., \& {Wilhelm}, K. 1997, \nat,
  386, 811

\bibitem[{{Judge} \& {Carlsson}(2010)}]{Judge+Carlsson2010}
{Judge}, P.~G. \& {Carlsson}, M. 2010, \apj, 719, 469

\bibitem[{{Kim} {et~al.}(2008){Kim}, {Bong}, {Park}, {Cho}, {Moon}, \&
  {Suematsu}}]{Kim+others2008}
{Kim}, Y., {Bong}, S., {Park}, Y., {et~al.} 2008, J. Korean Astron. Soc., 41,
  173

\bibitem[{{Kukhianidze} {et~al.}(2006){Kukhianidze}, {Zaqarashvili}, \&
  {Khutsishvili}}]{Kukhianidze+others2006}
{Kukhianidze}, V., {Zaqarashvili}, T.~V., \& {Khutsishvili}, E. 2006, \aap,
  449, L35

\bibitem[{{L{\'o}pez Ariste} \& {Casini}(2005)}]{LopezAriste+Casini2005}
{L{\'o}pez Ariste}, A. \& {Casini}, R. 2005, \aap, 436, 325

\bibitem[Maltby et al.(1986)]{Maltby+others1986} Maltby, P., Avrett, 
E.~H., Carlsson, M., Kjeldseth-Moe, O., Kurucz, R.~L., 
\& Loeser, R.\ 1986, \apj, 306, 284 



\bibitem[{{Musielak} \& {Ulmschneider}(2001)}]{Musielak+Ulmschneider2001}
{Musielak}, Z.~E. \& {Ulmschneider}, P. 2001, \aap, 370, 541

 
\bibitem[Nakariakov \& Verwichte(2005)]{Nakariakov+Verwichte2005} Nakariakov, V.~M., \& Verwichte, E.\ 2005, Living Reviews in Solar Physics, 2, 3 

\bibitem[{{Nakariakov} \& {Ofman}(2001)}]{Nakariakov+Ofman2001}
{Nakariakov}, V.~M. \& {Ofman}, L. 2001, \aap, 372, L53

\bibitem[{{Orozco Suarez} {et~al.}(2005){Orozco Suarez}, {Lagg}, \&
  {Solanki}}]{OrozcoSuarez+others2005}
{Orozco Suarez}, D., {Lagg}, A., \& {Solanki}, S.~K. 2005, in ESA Special
  Publication, Vol. 596, Chromospheric and Coronal Magnetic Fields, ed.
  {D.~E.~Innes, A.~Lagg, \& S.~A.~Solanki}

\bibitem[{{Pasachoff} {et~al.}(1968){Pasachoff}, {Noyes}, \&
  {Beckers}}]{Pasachoff+others1968}
{Pasachoff}, J.~M., {Noyes}, R.~W., \& {Beckers}, J.~M. 1968, \solphys, 5, 131

\bibitem[{{Roberts}(1979)}]{Roberts1979}
{Roberts}, B. 1979, \solphys, 61, 23


\bibitem[{{Rouppe van der Voort} {et~al.}(2009){Rouppe van der Voort},
  {Leenaarts}, {de Pontieu}, {Carlsson}, \&
  {Vissers}}]{RouppevanderVoort+others2009}
{Rouppe van der Voort}, L., {Leenaarts}, J., {de Pontieu}, B., {Carlsson}, M.,
  \& {Vissers}, G. 2009, \apj, 705, 272

\bibitem[{{S{\'a}nchez Almeida}(1997)}]{SanchezAlmeida1997}
{S{\'a}nchez Almeida}, J. 1997, \aap, 324, 763

\bibitem[{{Scharmer}(2006)}]{crisp}
{Scharmer}, G.~B. 2006, \aap, 447, 1111

\bibitem[{{Scharmer} {et~al.}(2003){Scharmer}, {Dettori}, {Lofdahl}, \&
  {Shand}}]{sstao}
{Scharmer}, G.~B., {Dettori}, P.~M., {Lofdahl}, M.~G., \& {Shand}, M. 2003, in
  Society of Photo-Optical Instrumentation Engineers (SPIE) Conference Series,
  Vol. 4853, , 370--380

\bibitem[{Selbing(2005)}]{Selbing2005}
Selbing, J. 2005, Master's thesis, Stockholm Observatory

\bibitem[{{Shoji} \& {Kurokawa}(1995)}]{Shoji+Kurokawa1995}
{Shoji}, M. \& {Kurokawa}, H. 1995, \pasj, 47, 239

\bibitem[{{Singh} \& {Dwivedi}(2007)}]{Singh+Dwivedi2007}
{Singh}, K.~A.~P. \& {Dwivedi}, B.~N. 2007, New Astron., 12, 479

\bibitem[{{Solanki} {et~al.}(2003){Solanki}, {Lagg}, {Woch}, {Krupp}, \&
  {Collados}}]{Solanki+others2003}
{Solanki}, S.~K., {Lagg}, A., {Woch}, J., {Krupp}, N., \& {Collados}, M. 2003,
  \nat, 425, 692

\bibitem[{{Spruit}(1981)}]{Spruit1981}
{Spruit}, H.~C. 1981, \aap, 98, 155

\bibitem[{{Terradas} {et~al.}(2011){Terradas}, {Andries}, \&
  {Verwichte}}]{Terradas+others2011}
{Terradas}, J., {Andries}, J., \& {Verwichte}, E. 2011, \aap, 527, A132

\bibitem[{{Trujillo Bueno} {et~al.}(2005){Trujillo Bueno}, {Merenda},
  {Centeno}, {Collados}, \& {Landi Degl'Innocenti}}]{TrujilloBueno+others2005}
{Trujillo Bueno}, J., {Merenda}, L., {Centeno}, R., {Collados}, M., \& {Landi
  Degl'Innocenti}, E. 2005, \apjl, 619, L191

\bibitem[{{Tsiropoula} {et~al.}(1994){Tsiropoula}, {Alissandrakis}, \&
  {Schmieder}}]{Tsiropoula+others1994}
{Tsiropoula}, G., {Alissandrakis}, C.~E., \& {Schmieder}, B. 1994, \aap, 290,
  285

\bibitem[van Doorsselaere et al.(2009)]{VanDoorsselaere+others2009} van 
Doorsselaere, T., Verwichte, E., \& Terradas, J.\ 2009, \ssr, 149, 299 



\bibitem[{{Van Doorsselaere} {et~al.}(2008){Van Doorsselaere}, {Nakariakov}, \&
  {Verwichte}}]{VanDoorsselaere+others2008}
{Van Doorsselaere}, T., {Nakariakov}, V.~M., \& {Verwichte}, E. 2008, \apjl,
  676, L73

\bibitem[{{van Noort} {et~al.}(2005){van Noort}, {Rouppe van der Voort}, \&
  {L{\"o}fdahl}}]{momfbd}
{van Noort}, M., {Rouppe van der Voort}, L., \& {L{\"o}fdahl}, M.~G. 2005,
  \solphys, 228, 191

\bibitem[{{Verwichte} {et~al.}(2010){Verwichte}, {Foullon}, \& {Van
  Doorsselaere}}]{Verwichte+others2010}
{Verwichte}, E., {Foullon}, C., \& {Van Doorsselaere}, T. 2010, \apj, 717, 458

\bibitem[{{Yi} {et~al.}(1992){Yi}, {Darvann}, \& {Molowny Horas}}]{destretch}
{Yi}, Z., {Darvann}, T.~A., \& {Molowny Horas}, R.~L. 1992, LEST Found.,
  Tech.~Rep., 56

\bibitem[{{Zaqarashvili} \& {Erd{\'e}lyi}(2009)}]{Zaqarashvili+Erdelyi2009}
{Zaqarashvili}, T.~V. \& {Erd{\'e}lyi}, R. 2009, \ssr, 149, 355

\bibitem[{{Zaqarashvili} {et~al.}(2007){Zaqarashvili}, {Khutsishvili},
  {Kukhianidze}, \& {Ramishvili}}]{Zaqarashvili+others2007}
{Zaqarashvili}, T.~V., {Khutsishvili}, E., {Kukhianidze}, V., \& {Ramishvili},
  G. 2007, \aap, 474, 627

\end{thebibliography}

\end{document}